\documentstyle[12pt]{article}
\input epsf

\begin{document}

\centerline{\large\bf Inflaton field potential producing
the exactly flat spectrum}
\centerline{\large\bf of adiabatic perturbations}

\medskip

\centerline{\rm Alexei A. Starobinsky}

\medskip

\centerline{\it Landau Institute for Theoretical Physics,
Russian Academy of Sciences,}
\centerline{\it 119334 Moscow, Russia}

\vspace{0.5cm}

\centerline{\bf Abstract}

\medskip
\noindent

Presented is the exact solution of the problem of finding a potential 
of an inflaton scalar field for which adiabatic perturbations 
generated during a de Sitter (inflationary) stage in the early 
Universe have the exactly flat (or, the Harrison-Zeldovich) initial 
spectrum. This solution lies outside the scope of the slow-roll 
approximation and higher-order corrections to it. The potential found 
depends on two arbitrary physical constants, one of those determines
an amplitude of perturbations. For small (zero) values of the other
constant, a long (infinite) inflationary stage with slow rolling of
the inflaton field exists. 

\medskip 


\vspace{0.5cm}

Great increase of amount and accuracy of cosmological observational 
data obtained already and expected in near future makes real the
ambitious programme of determination of an initial power spectrum 
$P_0(k)$ of inhomogeneous density perturbations in the Universe
directly from data. Then, if working within the scope of the simplest 
versions of the inflationary scenario of the early Universe (i.e.,
with one slowly rolling scalar field $\phi$ -- the inflaton), a 
natural next step is to reconstruct an effective self-interaction 
inflaton potential $V(\phi)$ leading to the generation of such
$P_0(k)$ from inflaton quantum vacuum fluctuations during a de Sitter 
(inflationary) stage. Note that in inflationary cosmology, an 
``initial''power spectrum means the spectrum that had been formed by 
the end of the inflationary stage, i.e. by the beginning of a 
post-inflationary power-law evolution of the Universe (it was first
calculated in \cite{MC81} for the $R+R^2$ model \cite{St80} and in
\cite{H82,St82,GP82} for the ``new'' inflationary model 
\cite{L82,AS82}). Here $k=|{\bf k}|$, and the spatial dependence 
$\exp(i{\bf kr})$ of all Fourier modes of adiabatic (scalar) 
perturbations is assumed. 

 From the mathematical point of view, since quantum generation of
perturbations reduces to a kind of scattering problem, the
reconstruction of $V(\phi)$ may be considered as a specific inverse 
scattering problem. Its solution is known in the slow-roll 
approximation for the inflaton field, including first {\cite{SL93},
second \cite{SG01} and third \cite{CGS04} order corrections to this 
basic approximation with respect to small slow-roll parameters, and 
in the so-called general slow-roll approximation \cite{DS02,JSGL05}. 
It is non-unique if additional information about the spectrum of
primordial gravitational waves generated during a de Sitter 
(inflationary) stage \cite{St79} is not used. Namely, generally there 
exists a one-parameter family of $V(\phi)$ producing the same 
$P_0(k)$ (here and below I don't count one more trivial free 
parameter corresponding to invariance with respect to a constant 
shift of the inflaton field: $\phi \to \phi +\phi_0$). However, an 
exact solution of the reconstruction problem is desirable for at 
least two purposes: \\
1) to investigate how good is the convergence of perturbation 
series in powers of slow-roll parameters (if it takes place at all); \\
2) to determine the exact degree of degeneracy of the problem,
i.e. to find the measure of a set of potentials producing the 
same perturbation spectrum. \\
In particular, the problem of accuracy of the slow-roll 
approximation prediction for $P_0(k)$ (including higher order 
corrections) has been intensively and critically studied recently 
using different methods: \cite{MS00}, \cite{HHH04} (the uniform 
approximation), \cite{CFLV05} (the improved WKB-approximation) and 
others.

By an exact solution I mean a solution of the following system of
equations for a spatially-flat Friedmann-Robertson-Walker (FRW)
background with a scale factor $a(t)$ and scalar (adiabatic)
perturbations described by the Mukhanov variable $Q\equiv u/a$:
\begin{eqnarray}
H^2 = {8\pi G\over 3}\left({\dot\phi^2\over 2} + V(\phi)\right)~, \\
\label{F-eq}
\ddot \phi + 3H\dot\phi + {dV\over d\phi}=0~, \\
\label{phi-eq}
{d^2u_k\over d\eta^2} + \left(k^2 - {1\over z}{d^2z\over d\eta^2}
\right)u_k = 0~,
\label{u-eq}
\end{eqnarray}
obtained without any approximations. Here
\begin{equation}
H = {\dot a\over a}~, ~~~z = {a\dot\phi\over H}~,~~~\eta = \int
{dt \over a(t)}~,
\end{equation}
dot means the derivative with respect to $t$, $u_k(\eta) 
\exp (i{\bf kr})$ is the wave function of a Fourier mode of the 
quantum field $u$ (the c-number multiplying the Fock annihilation
operator $\hat a_{\bf k}$), and $c=\hbar =1$ is put throughout the 
paper. The variable $Q$ \cite{M85} is equal to $\delta \phi_L + 
{\dot \phi\over H}\Phi$ in the longitudinal gauge ($\Phi$ is the 
quasi-Newtonian gravitational potential), or to $\delta\phi_S - 
{\dot \phi \over 6H}(\mu + \lambda)$ in the synchronous gauge 
($\mu$ and $\lambda$ are the Lifshits variables). The normalized 
initial condition for $u_k$ corresponding to the adiabatic vacuum 
at $t\to -\infty$ ($\eta \to - \infty$) is
\begin{equation}
u_k = {e^{-ik\eta}\over \sqrt{2k}}~.
\label{incon}
\end{equation}
At late times during an inflationary stage in the super-horizon 
regime ($k\ll aH,~\eta\to 0$), 
\begin{equation}
{u_k\over z}={HQ_k\over \dot\phi} \to const = \zeta(k)
\end{equation} 
($\zeta = -h/2$ in the notation of \cite{St82}).  

Then the initial spectrum of adiabatic perturbations for a
post-inflationary cosmology in the super-horizon regime is
(assuming the absence of non-diagonal pressure components):
\begin{equation}
<\Phi^2>=\left(1-{H\over a}\int_0^t a\,dt\right)^2<\zeta^2>=
\left(1-{H\over a}\int_0^t a\,dt\right)^2\int P_0(k)\,{dk\over k}~, 
~~P_0(k)= {k^3\zeta^2(k)\over 2\pi^2}~.
\end{equation}
Here $t=0$ corresponds to the end of inflation. For historical
reasons, the slope $n_S$ of the spectrum is defined with respect
to density perturbations in the non-relativistic dark matter + 
baryon component at the present time, $(\delta \rho)_k = - k^2
\Phi_k/4\pi Ga^2$ before integration over $d^3k$. So, 
$n_S = 1 + {d\ln P_0(k)\over d\ln k}$. Finally, using the equation
$\dot H = -4\pi G \dot\phi^2$ that follows form Eqs. (\ref{F-eq})
and (\ref{phi-eq}), Eq. (\ref{F-eq}) can be recast in the 
Hamilton-Jacobi form \cite{MSB90}:
\begin{equation}
H^2(\phi) - {H'^2(\phi)\over 12\pi G} = {8\pi G\over 3}V(\phi)~,
\label{HJ-eq}
\end{equation}
where the prime denotes the derivative with respect to $\phi$.

Exact solutions of the inverse problem of reconstruction of 
$V(\phi)$ given $P_0(k)$ are known for the following two cases
only, if not speaking about solutions describing universes  
collapsing towards a singularity. \\
1) A power-law perturbation spectrum with the slope 
$n_S = const < 1$~\cite{LM85}. Then 
\begin{equation}
V(\phi)\propto H^2(\phi) \propto \exp \left({\pm \sqrt{{16\pi 
G\over q}}\phi}\right)~,~~a(t)\propto t^q~,~~q={3-n_S\over 
1-n_S}>1~.
\label{pow}
\end{equation}  
This is just the power-law inflation. Considered as a function of 
$\phi (t)$, $H$ is related to $V(\phi)$ through Eq.~(\ref{HJ-eq}).
Note, however, that this is not the only potential producing the 
$n_S=const<1$ spectrum.\\
2) The case when no perturbations are generated at all (no
real created quanta of the inflaton field)~\cite{E96}:
\begin{equation}
H(\phi)=H_1\exp (2\pi G \phi^2)~,~~V(\phi)={3H_1^2\over
8\pi G} \left(1-{4\pi G\phi^2\over 3}\right)\exp (4\pi G \phi^2)~.
\label{no}
\end{equation}
In literature, this case is sometimes incorrectly referred as the 
potential generating the $n_S=3$ perturbation spectrum. However, 
one should not forget that generated perturbations are quantum (even
quantum-gravitational) and require renormalization. After subtraction 
of the vacuum energy $\omega (t)/2=k/2a(t)$ of each mode, no created 
fluctuations remain in this case. Moreover, a number of real
inflaton quanta generated in each perturbation mode ${\bf k}$
should be large, because in the opposite case they may not be
interpreted as classical perturbations after the end of inflation
(see~\cite{PS96} for a more detailed discussion of this point).

Strictly speaking, there is no exit from inflation for the potential
(\ref{pow}), and the potential (\ref{no}) does not admit a low 
curvature regime at all. However, in the former case $V(\phi)$ can
be deformed such that it reaches zero at a sufficiently large value of
$\phi$. This will result in a very small change of the perturbation
spectrum at present scales of interest that may be safely neglected.
Sometimes, the case of a parabolic potential near its maximum $V(\phi)=
V_0 - {m^2\phi^2\over 2}$ is mentioned as an exactly soluble case.
However, it is not such the one in our terminology since in this case
$H(\phi)$ is approximated by the constant value $H_0=
\sqrt{8\pi GV_0/3}$.

In this paper, a family of exact solutions for the case $n_S = 1$
is constructed. It is just the initial spectrum proposed by Harrison
and Zeldovich \cite{HZ}, after all, for beauty reasons. Note that it
satisfies the most recent CMB data \cite{Arch,WMAP}. Let us first 
consider what follows for this case from the slow-roll approximation.
Then, the leading term in the power spectrum reads
\begin{equation}
k^3\zeta^2(k)\propto \left({V^3\over V'^2}\right)_{t=t_k}~,
\end{equation}
where $t_k$ is the moment when $k=aH$. It is clear that, to get
$n_S=1$, $V^{3/2}/V'$ should not depend on $\phi$. Therefore,
$V(\phi)\propto \phi^{-2}$. Note that this solution of the 
reconstruction problem is unique for a given amplitude of the flat 
spectrum. This kind of inflation was dubbed intermediate inflation in 
\cite{BL93} (see also \cite{R05}). Its scale factor behaviour is $a(t) 
\propto \exp \left(const\cdot t^{2/3}\right)$. Once more, it does not
have an exit from inflation, so it should be modified at large
$\phi$. A next order slow-roll correction to this potential was 
considered in \cite{VCK04}.   

To obtain an exact solution for $H(\phi)$ and $V(\phi)$ in the case 
$n_S=1$, note first that, for 
\begin{equation}
{1\over z}\,{d^2z\over d\eta^2} = {2\over \eta^2}~,
\label{z-term}
\end{equation} 
Eq. (\ref{u-eq}) reduces to the equation for a massless scalar field
in the de Sitter background and has the solution
\begin{equation}
u_k = {e^{-ik\eta}\over \sqrt{2k}}\left(1 - {i\over k\eta}\right)
\end {equation}
satisfying the initial condition (\ref{incon}). Let us write the 
general solution of Eq. (\ref{z-term}) in the form
\begin{equation}
z = {B\over |\eta|}\left(1 + {|\eta|^3\over \eta_0^3}\right), 
~~\eta < 0~,
\label{z}
\end{equation}
where $A,~\eta_0$ are constants. The limiting case $\eta_0 \to 0$,
when the first term in brackets may be neglected, is not interesting 
because it corresponds to a collapsing universe (however, it is 
``dual'' to the case $\eta_0\to\infty$ considered below). The power
spectrum of the growing perturbation mode is $P_0(k)=1/4\pi^2B^2$ and
does not depend on $\eta_0$ ($\eta_0$ appears in the amplitude of the
decaying mode only and makes it non-scale-free). Thus, we have got the
exactly flat spectrum. Present observational CMB data \cite{WMAP} fix 
the quantity $B$ with $\approx 10\%$ accuracy:
\begin{equation}
{1\over 2\pi B}= 4.8\cdot 10^{-5}\left({A\over 0.9}\exp(\tau - 0.17)
\right)^{1/2},
\label{B}
\end{equation}
where $A$ is the quantity introduced in \cite{WMAP} and $\tau$ is the 
optical length after recombination. In this notation, $A=0.9$ 
corresponds to the value $A=4.3\cdot 10^{-4}$ of the other quantity
$A$ introduced in \cite{St83} to characterize an amplitude of initial 
perturbations (and conjectured to lie in the range $(3-10)\cdot
10^{-4}$ in that paper).    

Since the aim of this paper is to find {\em some} exact solution, 
I will not investigate if there exist other forms of $z$ leading to 
the $n_S=1$ spectrum, too. The absence of other solutions for $z$
would immediately follow from scaling arguments if we assume that 
$u_k\propto k^{-1/2}f(k\eta)$ for {\em all} $\eta$. However, the 
latter assumption might not be necessary. Moreover, I will consider
only one particular case of Eq. (\ref{z}) corresponding to the limit
$\eta_0\to \infty$.

So, let $z=-B/\eta$. Let us express all quantities of interest as 
functions of $\phi$:
\begin{eqnarray}
t= -4\pi G\int {d\phi\over H'}~, ~~\ln a = \int H(t)\, dt =
-4\pi G\int {H\over H'}\,d\phi~, \nonumber \\
\eta = \int {dt \over a(t)}= -4\pi G\int {d\phi\over H'}\,\exp\left(
4\pi G\int{H\over H'}\,d\phi\right)~, \label {expr} \\
z={a\dot\phi\over H}= -{H'\over 4\pi GH}\,\exp\left(-4\pi G\int
{H\over H'}\,d\phi\right)~. \nonumber
\end{eqnarray}
Equating the last line in Eq. (\ref{expr}) to $-B/\eta$, we get the
following equation:
\begin{equation}
\int P(\phi)\,d\phi = -BHP~, ~~P\equiv {4\pi G\over H'}\exp\left(
4\pi G\int {H\over H'}\,d\phi\right)~.
\label{P}
\end{equation}
After differentiation, Eq. (\ref{P}) reduces to $P=-B(HP'+H'P)$,
or
\begin{equation}
{4\pi GH^2\over H'}-{HH''\over H'}+H'+{1\over B}=0~.
\label{finaleq}
\end{equation}
Let us introduce dimensionless variables 
\begin{equation}
x=\sqrt{4\pi G}\phi~, ~~y = B\sqrt{4\pi G}H~, ~~v(x)={32\pi^2G^2B^2
\over 3}V(\phi)~.
\end{equation}
Then, from (\ref{HJ-eq}), $v = y^2 - (1/3)(dy/dx)^2$. For these 
variables, Eq. (\ref{finaleq}) reads:
\begin{equation}
y\,{d^2y\over dx^2}= \left({dy\over dx}\right)^2 + {dy\over dx} + y^2~.
\end{equation} 
After dividing by $y^2$, the last equation can be integrated to
$dy/dx = xy-1$ (an integration constant is excluded by shifting $x$, 
i.e., $\phi$). Therefore,
\begin{equation}
y=e^{x^2/2}\left(\int_x^{\infty}e^{-\tilde x^2/2}\,d\tilde x +
C\right)~,
\label{answ}
\end{equation}
where $C$ is another integration constant. This just yields us a 
one-parameter family of solutions having $n_S=1$. The so-called
slow-roll parameters for this solution:
\begin{eqnarray}
\epsilon(\phi)\equiv {1\over 4\pi G}\,{H'^2\over H^2}=\left({1\over y}
- x\right)^2, \nonumber \\
\tilde\eta(\phi)\equiv {1\over 4\pi G}\,{H''\over H}={1\over y}\,{d^2y
\over dx^2}=x^2-{x\over y}+1~.
\label{srp}
\end{eqnarray}

The partial solution with $C=0$ has an infinite inflationary stage
which is just described by the slow-roll approximation for $x\gg
1$. Its graph is plotted in Fig.1. Its large-$x$ expansion is
\begin{equation}
y={1\over x}-{1\over x^3}+{3\over x^5}-{15\over x^7}+...,~~
v={1\over x^2}-{7\over 3x^4} +{9\over x^6} - ...
\end{equation}
It is straightforward to check that it leads to $n_S=1$ (as it should 
be) for the first \cite{SL93} and second \cite{SG01} order corrections 
to the slow-roll approximation. However, these corrections miss the
whole 1-parametric family with $C\not= 0$ completely.

\begin{figure}
\epsfbox{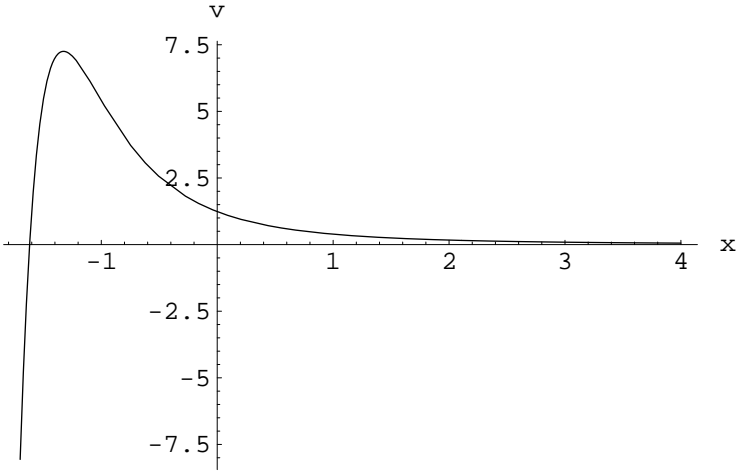}
\caption{The dimensionless potential $v(x)$ for the $C=0$ case.}
\end{figure}

 For $x<0$, the solution with $C=0$ has rather peculiar behaviour: the
potential $v(x)$ reaches the maximum value $v_{max}\approx 7.252$ at
$x\approx -1.326$, becomes zero at $x\approx -1.618$ and then going
to $-\infty$ at $x\to -\infty$ (however, such effective potentials are
considered in string inspired models now). In the latter limit, $y\to
\infty$, so we get an initial curvature singularity at a finite 
proper time $t_0<0$. If $t=0$ is the moment when $x=0~(v(0)=
{\pi\over 2}-{1\over 3}\approx 1.237)$ and the inflationary stage
begins, then $|t_0|\sim H^{-1}(0)\sim BG^{1/2}$. The scale factor
reaches zero very slowly: $a(t)\propto |\ln(t-t_0)|^{-1/2}$ for
$t\to t_0$. Still the Riemann tensor is not twice integrable for 
$t\to t_0$, so this singularity is a strong one. The same refers to 
all initially expanding ($y>0$) solutions with $C\not= 0$ and 
$C>-\sqrt{2\pi}$ -- they all begin from such a singularity.

By taking $C<0$ and very small, it becomes possible to construct a
solution with a long but finite inflationary stage. Namely, if 
$C=-\sqrt{3}x_1^{-2}\exp(-x_1^2/2)$ with $x_1\gg 1$, then $v(x)$ 
becomes zero at $x=x_1$ ($y$ still remains $\sim x_1^{-1}$). In this
case inflation ends ($\epsilon,~|\tilde \eta|\sim 1$) at $x=x_1-
{\cal O}(x_1^{-1})$. The total number of e-folds is $N_{tot}=
2\pi G\phi_1^2 = x_1^2/2$. Thus, $C\sim \exp(-N_{tot})$ that is in
agreement with the general principle that terms not caught by an
arbitrary order of a WKB-type expansion are exponentially small.
For $x\ge x_1$, one may put $v\equiv 0$. Then the kinetic dominated
phase $a(t)\propto t^{1/3}$ follows the inflationary stage. Or, we
may assume that $v$ has a local minimum $v={1\over 2}\,\mu^2 
(x-x_1)^2$ around this point. It results in oscillations in $\phi$ 
and the matter-dominated post-inflationary stage $a(t)\propto 
t^{2/3}$.

Finally, note that the spectrum of gravitational waves (GW) is not 
flat for this model: for $1\ll x \ll x_1$, the tensor-scalar ratio 
and the slope of the GW initial power spectrum $r=-8n_T=16/x^2 = 8/N$
where $N$ is the number of e-folds from the {\em beginning} of
inflation. The present upper observational bound $r<0.36$ 
\cite{SMD05} requires $N>22$ for the comoving scale crossing the 
Hubble radius at present. So, $N_{tot}$ should exceed $\sim 70$ in 
this model.

The research was partially supported by the Russian Foundation for 
Fundamental Research, grant No. 05-02-17450, by the Research Programme 
``Elementary Particles'' of the Russian Academy of Sciences and by the 
scientific school grant No. 2338.2003.2 of the Russian Ministry of 
Education and Science.

\vfill
\end{document}